\documentclass{kapproc} 

\usepackage{procps} 

\usepackage[dvips]{graphicx}

\upperandlowercase

\kluwerbib 

\begin{document}

\articletitle{Optical Studies of Isolated Neutron Stars and Their Environments}

\author{Roberto P. Mignani}
\affil{European Southern Observatory}
\email{rmignani@eso.org}

\begin{abstract} 
The results  of optical studies  of  Isolated Neutron Stars
(INSs), their Pulsar-Wind Nebulae (PWNe) and Pulsar Bow Shocks are
reviewed and discussed. 
\end{abstract}

\begin{keywords}
Pulsars, Isolated Neutron Stars, Optical
\end{keywords}

\section{Isolated Neutron Stars}

\subsection{The identification record}
\begin{table}
\small
\begin{tabular}{l|lll|lll|l} \hline
{\em Name} & {\em Year}  & {\em Tel} & {\em Size} & {\em mag} & {\em d(kpc)$^{*}$} & {\em $A_{V}$}  &    {\em Identification} \\  \hline
Crab       	   	&1969	&Steward &0.9m    &16.6	&1.73	&1.6	&Pulsations               \\
Vela              	&1976	&CTIO	 &4m      &23.6	&0.23	&0.2 	&Pulsations               \\
B0540-69           	&1984	&CTIO 	 &4m      &22  	&49.4	&0.6	&Pulsations	          \\
Geminga         	&1987	&CFHT	 &3.6m    &25.5	&0.16	&0.07	&Proper Motion/Pulsations \\
B0656+14        	&1994	&NTT     &3.5m    &25  	&0.29	&0.09	&Pulsations/Proper Motion \\
B0950+08           	&1996	&HST     &2.4m    &27.1	&0.26	&0.03	&Position/Photometry	  \\
B1929+10            	&1996	&HST	 &2.4m    &25.6	&0.33	&0.15	&Proper Motion		  \\
B1055-52           	&1997	&HST	 &2.4m    &24.9	&0.72	&0.22	&Position                 \\
RXJ1856-3754        	&1997	&HST	 &2.4m    &25.7	&0.14	&0.12  	&Proper Motion 		  \\    
J0720-3125          	&1998	&Keck	 &10m     &26.7	&    	&0.30 	&Proper Motion  	  \\
B1509-58           	&2000	&VLT     &8.2m    &25.7&4.18	&5.2	&Position		  \\
RXJ1308.6+2127     	&2002	&HST     &2.4m    &28.6	&    	&0.14	&Position		  \\
RXJ1605.3+3249    	&2003	&HST	 &2.4m    &26.8	&    	&0.06	&Proper Motion 		  \\
J0437-4715        	&2004	&HST	 &2.4m    &    	&0.14	&0.11  	&Spectroscopy		   \\\hline
\end{tabular}

$^{*}$http://rsd-www.nrl.navy.mil/7213/lazio/ne\_model/

\caption{INSs identification  status.  The columns give  the name, the
year  of  the proposed  identification,  the  used  telescope and  its
aperture, the  magnitude ($V$-band when available),  the distance, the
interstellar absorption $A_V$ and the identification evidence.  }
\end{table}

\noindent
The  Isolated Neutron Stars  (INSs) with an  associated optical
counterpart, including both rotation-powered pulsars and the so-called
X-ray  Dim INSs  (XDINSs), are  now 14  (Table1; see  also  Mignani et
al. 2004a) i.e.  about as many as those detected in X-rays in
the  pre-ROSAT era.   Mostly thanks  to the  HST and the  high UV
sensitivity of the FOC and  the STIS, the initial identification score
of two  objects per decade  has increased  to almost {\it  one per
year}.  As a matter of fact, in the last 10 years HST has detected all
the INSs it  was targetted to.  On the  other hand, large ground-based
telescopes like  the Kecks and the  VLT so far played  only a marginal
role.   Recent  HST  observations   might  have  identified  also  the
counterpart of the  young 16 ms pulsar PSR  J0537-6910 (Mignani et al.
2004b).   Owing to  their  intrinsic faintness,  most  INSs have  been
detected only  because of their close distance  and small interstellar
absorption.   This made  also  possible  to use  proper  motion as  an
alternative,  and indeed very  efficient, identification  technique to
optical timing.

\begin{table}[h]
\small
\begin{tabular}{l|ll|lll|l} \hline
{\em Name} &  {\em Spec.(\AA)}  & {\em Phot.} & {\em $\alpha$} & {\em T} & {\em Comments}  & {\em Pol.$^{**}$} \\ \hline
Crab  		& 1100-9000 &UV,UBVRI,JHK& -0.11 & - & $PL_{o} <PL_{x}$    & 20\%  (IP) \\
                &           &            &       &   &   & 40\% (OP) \\
B1509-58  	&           &R           &       &   & 		            & 10\%  (TI)\\ 
B0540-69 	& 2500-5500 &UBVRI       & +0.2  &-  & $PL_{o} < PL_{x}$    & 5 \%  (TI)\\ 
Vela    	& 4500-8600 &UV,UBVRI,JH & +0.12 &-  & $PL_{o} \sim PL_{x}$ & 8.5\% (TI) \\ \hline 
B0656+14 	&           &UV,UBVRI,JHK& +0.45 &8.5& $PL_{o} \sim PL_{x}$ & 100\% (IP) \\ 
        	&           &            &       &   & $BB_{o} \sim BB_{x}$ &     \\ 
Geminga 	& 3700-8000 &UV,UBVRI,JH & +0.8  &4.5& $PL_{o} < PL_{x}$    &     \\
        	&           &            &       &   & $BB_{o} \sim BB_{x}$ &     \\
B1055-52 	&           &U           &       &   &                      &     \\ \hline
B1929+10 	&           &UV,U        & +0.5  &-  & $PL_{o} < PL_{x}$    &     \\ 
B0950+08		&           &U,BVI       & +0.65 &-  & $PL_{o} \sim PL_{x}$   &     \\ \hline 
J0437-4715	&1150-1700  &            & -     &1.0& $BB_{o} >
BB_{x}$    &     \\ \hline 
RXJ0720-3125   	&           & UV,UBVR    & -1.4  & 4 & $BB_{o} > BB_{x}$    &     \\ 
RXJ1856-3754    &3600-9000  &UV,UBV      & -     &2.3& $BB_{o} > BB_{x}$    &     \\ 
RXJ1605.3+3249	&           &VR          &       &   & $BB_{o} > BB_{x}$    &     \\
RXJ1308.6+2127	&           &V           &       &   & $BB_{o} > BB_{x}$    &     \\ \hline
\end{tabular}

$^{**}$IP=Inter Pulse; OP=Off Pulse; TI=Time Integrated

\caption{Optical INS database grouped by age decades. The columns give
the name,  spectroscopy, photometry,  the spectral index  $\alpha$ and
temperature (in  units of $10^{5}  K$) of the power-law  and blackbody
components  ($PL_{o}$;$BB_{o}$) and  the comparison  with  the optical
extrapolation  of the  X-rays  ones ($PL_{x}$;$BB_{x}$).  Polarization
measures are indicated in the last column.}
\end{table}

\subsection{Photometry and Spectroscopy}

\noindent
Table 2 summarizes the optical  INSs database (see also Mignani et al.
 2004a  and references  therein).   Only for  six  of them  optical/UV
 spectroscopy is available and only  for four photometry spans all the
 way from the  IR to the UV.  This is crucial  to identify thermal and
 non-thermal spectral  components whose contributions  are expected to
 be markedly different in the IR and in the UV.
As a general trend, the spectrum grows in complexity with the age from
 a single power-law  (PL) dominated to a composite  one featuring both
 PL and blackbody (BB) components.
While  in  some  cases  the  optical PL/BB  components  do  match  the
 extrapolation of  the X-ray  ones, apparently this  is not  a general
 rule, which  suggests that the optical and  X-ray emission mechanisms
 are not always  related to each other.  In  particular, for XDINs the
 optical  BB   spectrum  appears   to  be  systematically   above  the
 extrapolation  of the  X-ray  one.  Apart from  the  decrease of  the
 temperature of  the BB component,  which follows from the  cooling of
 the  neutron  star surface,  there  is  no  clear indication  for  an
 evolution of the spectral parameters with the neutron star's age.

\subsection{Timing}

\noindent
 After the ``historical'' optical pulsars Crab, Vela and PSR B0540-69,
pulsations have been clearly  detected from Geminga (Romani \& Pavlov,
in preparation) and PSR B0656+14  (Gull et al., 2004) thanks to recent
HST/STIS  observations, confirming  and improving  the  earlier 
results of Shearer  et al.  (1997) and Shearer et  al.  (1998). In all
cases but PSR B0540-60 the  lightcurves are double-peaked and for both
the  Crab  and  Geminga  the  peaks  are aligned  in  phase  with  the
$\gamma$-ray ones.

\subsection{Polarimetry}

\noindent
Till recently, the only INS with measured optical polarization was the
Crab (see  also Kamback et  al., these proceedings).   The breacktrough
came with the VLT  which measured the time-integrated polarization for
PSR  B0540-69, Vela  and PSR  B1509-58 (Wagner  \& Seifert  2000). More
recently,  time-resolved polarization  was measured  for  PSR B0656+14
(Kern et al. 2003).

\section{Pulsar-Wind Nebulae}

\noindent
So far, Pulsar  Wind Nebulae (PWNe) have been  detected in the optical
only for two  young pulsars: the Crab (e.g., Hester  et al.  2002) and
PSR B0540-69 (Caraveo et al.   2000).  In both cases, HST observations
have  clearly  resolved  the  counterparts  of  the  X-ray  structures
detected by {\sl Chandra}.  The  PWN around the Vela pulsar originally
claimed  by  \"Ogelman  et  al.   (1989)  was  not  confirmed  by  HST
observations (Mignani et al. 2003) which put 3$\sigma$ upper limits of
$\approx  27.9$  and $\approx  28.3$--27.8  mag  arcsec$^{-2}$ on  the
brightness of the inner and  outer X-ray PWN, respectively, i.e. close
the  extrapolation of  the  X-ray/radio data.   Comparable deep  upper
limits  were  set  on the  optical  emission  of  the PWN  around  PSR
J0537-6910 (Mignani et al. in preparation).

\section{Pulsar Bow-Shocks}

\noindent
The  interaction between the  pulsar's relativistic  wind and  the ISM
compressed by  the pulsar supersonic  motion originates a  shock which
ionizes the ISM and produce emission in $H_{\alpha}$. The $H_{\alpha}$
luminosity  ($L_{H_{\alpha}}$)  depends  on  the  pulsar's  rotational
energy loss ($\dot E$), on  the neutron star's velocity ($v_{NS}$) and
on the  fraction $X$  of neutral Hydrogen  in the ISM.   The bow-shock
shape  tends to  be symmetric  wrt  the pulsar  proper motion  ($\mu$)
direction,  with  deviations  determined  by  the  local  ISM  density
distribution,  featuring  either  arc-like or  bullet-like  structures
according to  the perspective.  So  far, optical bow-shocks  have been
clearly identified around 6 INSs  (see Table 3).  By imposing pressure
balance between  the (radial) pulsar wind  and the ISM  and assuming a
geometrical model for the bow-shock
one can derive  the local ISM density ($\rho_{ISM}$)  and the angle of
the  pulsar's velocity vector  wrt the  line of  sight.  From  the 3-D
velocity,  the  distance and  the  age one  can  then  trace back  the
galactic  orbital motion  of the  pulsar, given  a  galactic potential
model,  and localize its  birth place,  hence identify  its progenitor
stellar population.

\begin{table}[h]
\small
\begin{tabular}{l|ll|lll|l|l} \hline
{\em Name}   & {\em Log(Age)} & {\em Log($\dot E$)}      & {\em $\mu$} & {\em d}  & {\em $v_{NS}$} & {\em Log($L_{Ha}$)}  &{\em Comment} \\  
             & (yrs)          &  & (mas/yr)   & (kpc)    & (km/s)  &  & \\ \hline
B0740-28     &5.20	&35.14   &29	&1.9  &204  & $\sim$29 &\\
B2224+65     &6.05	&33.08	 &182	&2    &1700 & $\sim$30 &\\
B1957+20     &9.18	&35.20	 &30.4	&1.53 &225  & $\sim$31 &ms,binary\\
J0437-4715   &9.2	&34.07	 &141	&0.14 &98   & $\sim$28 &ms,binary\\
J2124-3358   &9.8	&33.63   &52.6	&0.27 &72   & $\sim$27 &ms,isolated\\
RXJ1856-3754 &?	        &?       &333	&0.14 &220  & $\sim$26 &\\ \hline
\end{tabular}
\caption{INSs with $H_{\alpha}$ bow-shocks. Luminosities and $\dot E$
are in units of erg~cm$^{-2}$~s$^{-1}$.} 
\end{table}

%

\end{document}